\documentclass[a4paper]{spicawStyle}
\usepackage{graphicx,natbib}

\begin{document}

\title{Simulated [{\sc Cii}] observations for SPICA/SAFARI}

\author{F. Levrier\inst{1} \and M. Gerin\inst{1} \and P. Hennebelle\inst{1} \and E. Falgarone\inst{1} \and F. Le Petit\inst{2} \and J.R. Goicoechea\inst{3}} 

\institute{
LERMA/LRA - UMR 8112 - ENS Paris, 24 rue Lhomond 75231 Paris CEDEX 05, France
\and 
LUTH - UMR 8102 - Observatoire de Paris-Meudon, 5 place Jules Janssen, 92195 Meudon CEDEX, France
\and
Centro de Astrobiolog{\'i}a, CSIC-INTA - Madrid, Spain
}

\maketitle 

\begin{abstract}

We investigate the case of [{\sc Cii}] 158$\mu$m observations for SPICA/SAFARI using a three-dimensional magnetohydrodynamical (MHD) simulation of the diffuse interstellar medium (ISM) and the Meudon PDR code. The MHD simulation consists of two converging flows of warm gas ($10^4$ K) within a cubic box 50 pc in length. The interplay of thermal instability, magnetic field and self-gravity leads to the formation of cold, dense clumps within a warm, turbulent interclump medium. We sample several clumps along a line of sight through the simulated cube and use them as input density profiles in the Meudon PDR code. This allows us to derive intensity predictions for the [{\sc Cii}] 158$\mu$m line and provide time estimates for the mapping of a given sky area.

\end{abstract}

\section{Introduction}

The fine structure line of ionised carbon [{\sc Cii}] at 157.7~$\mu$m is one of
the most prominent far infrared (FIR) line. It is widespread in the Milky Way as
revealed by the COBE-FIRAS observations, and detected in the diffuse ISM \citep{levrierf:I02}, photo-dissociation regions as well as local and distant
galaxies \citep[and references therein]{levrierf:M09}. [{\sc Cii}] is one of the main cooling lines of the neutral ISM, in
all environments where carbon
is mostly ionised, that is where UV-photons with $h\nu \geq 11.3~\mathrm{eV}$ are available. Unfortunately, the present information on the spatial
structure of the [{\sc Cii}] emission is very limited, because of the small
telescope size of previous far infrared missions. While Herschel and SOFIA
will improve the situation for bright regions, their sensitivity will be
limited for the extended emission from either the diffuse interstellar
medium, or the outskirts of molecular clouds.
The structure of the [{\sc Cii}] emission is expected to reveal the underlying
structure of the matter, and provide information on the mechanisms
responsible for the fragmentation. In this paper, we explore the
perspectives offered by SPICA for
mapping the interstellar medium.

\section{The MHD simulation}

To model the turbulent interstellar medium, we used a data cube from an MHD simulation performed by \citet{levrierf:H08} with the RAMSES code \citep{levrierf:T02,levrierf:FHT06}, which makes use of adaptive mesh refinement (AMR) methods. The cube is 50 parsecs on each side, with potentially 14 levels of refinement, leading to a maximal resolution of 3 mpc or 630 AU. However, to simplify the analysis, we used a regularly gridded (1024$^3$) version of the cube, with a pixel size $\sim$0.05 pc.

In the initial setup of the simulation, two converging flows of warm gas ($8000$ K) are entering the cube along the $X$ axis from two opposite sides. The flows' velocity is twice the sound speed of the warm neutral medium (WNM), with transverse and longitudinal modulations of 100\% amplitude with a 10pc periodicity. Periodic boundary conditions are imposed on the remaining four sides. The simulation includes a consistent treatment of hydrodynamics, magnetic fields, atomic cooling processes and self-gravity, and therefore implies some heavy computing ($\sim$30,000 CPU hours).

\begin{figure}[ht]
  \begin{center}
    \includegraphics[width=8.7 cm]{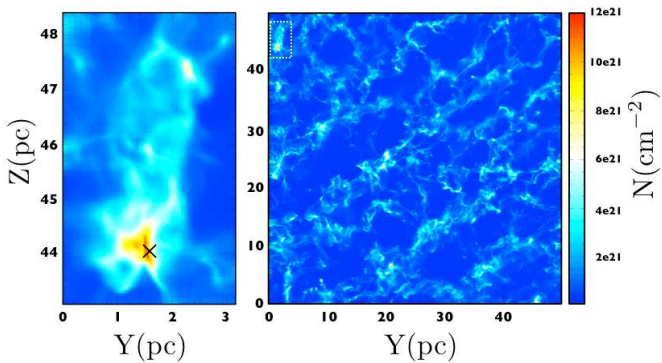}
  \end{center}
  \caption{Total gas column density along the $X$ direction - that of the converging inflows - in the MHD simulation. The figure on the right shows the entire 50 {\rm pc} $\times$ 50 {\rm pc} field. The figure on the left presents a zoom on a clump at the top right corner of the field. The line of sight used in the analysis is marked with a black cross.}
\label{levrierf_fig:fig1}
\end{figure}

Figure \ref{levrierf_fig:fig1} shows the total gas column density in the cube when viewed along the $X$ direction, which we take as the observer's direction to ensure statistical homogeneity of the observed field. Our chosen line of sight is displayed, and passes through one of the cold dense clumps formed by thermal instability in the turbulent flow. These clumps ($n\sim 10^3$ cm$^{-3}$ and $T\sim 10$ K) are found along filamentary structures amidst a much more diffuse and hot interclump medium ($n\sim 1$ cm$^{-3}$ and $T\sim 10^4$ K).

\begin{figure}[ht]
  \begin{center}
    \includegraphics[width=8.7 cm]{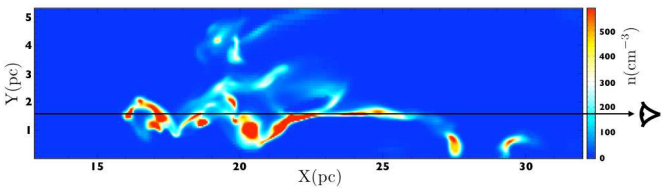}
  \end{center}
  \caption{$X-Y$ total gas density cut through the clump of interest. The line of sight used in the analysis is indicated, and the observer is located at the right side of the figure.}  
\label{levrierf_fig:fig2}
\end{figure}

Figure \ref{levrierf_fig:fig2} shows a total gas density cut through the cube in the $X-Y$ plane containing the line of sight. Complex density structures along this line appear quite clearly.

When it comes to simulating the observation of this field, we shall need to specify the angular pixel size. To match the SAFARI specifications, let us consider that the cloud is located 1.75 kpc away. It is therefore 1.6$^\circ$ across, and the pixel size is 5.75", i.e. that of the SAFARI pixels. 

\section{PDR code}
 
The Meudon PDR code \citep{levrierf:JLB93,levrierf:FLP06} is a publicly available model\footnote{\tt http://pdr.obspm.fr} aimed at describing the atomic and molecular structure of interstellar clouds, especially photon-dominated regions (PDR). It considers a plane-parallel slab of gas and dust illuminated on either or both sides by an incoming radiation field which can be the interstellar standard radiation field (ISRF) or the light from a nearby star. The density profile within the slab can either be constant, computed from an isobaric constraint or user-defined. Radiative transfer in the UV is solved at each point in the cloud \citep{levrierf:GLB07}, taking into account continuum absorption by dust grains and molecular absorption lines. Thermal balance is computed via the inclusion of a number of heating (photoelectric effect, cosmic rays) and cooling processes (infrared and millimeter emission lines) and an extensive chemical network. With abundances and level populations of the most prominent species available, line intensities and column densities can be obtained.

\begin{figure}[ht]
  \begin{center}
    \includegraphics[width=8.7 cm]{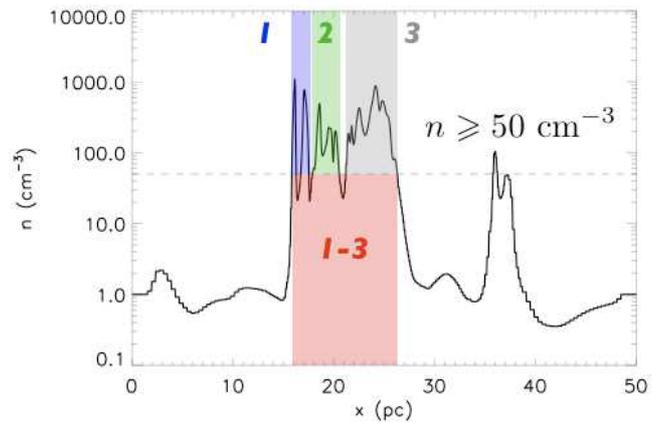}
  \end{center}
  \caption{Total gas density profile along the line of sight. Regions where density exceeds 50 particles per cubic centimeter are designated as "clumps" and numbered from 1 to 3. The whole region marked '1-3' is designated as "cloud" in the text.}  
\label{levrierf_fig:fig3}
\end{figure}

Convergence issues make it difficult to run the code reliably on the more diffuse regions. Consequently, we opted not to use the entire line of sight as input density profile in the PDR code, but rather extracted dense "clumps" (see Fig. \ref{levrierf_fig:fig3}) using a threshold value of 50 cm$^{-3}$ - which is about the critical density for C$^+$ in electronic collisions at $T_e=10^4$ K - and ran the PDR code on these clumps separately.

The one-dimensional geometry imposed by the code makes shielding an important point to remember when considering results. The complexity of density structures in the simulation cube implies that the radiation field should be higher in the interior regions than expected from the code, as it can be expected that the photons penetrate deep into the clouds, and, consequently, that the addition of results from separate clumps might give a better idea of the output emission.

To estimate this effect, we also ran the PDR code on the entire "cloud" marked '1-3' on Fig. \ref{levrierf_fig:fig3}, so in total four times. The density in the small interclump regions does not fall to such low values that the code may not converge.

\section{Results}
 
\begin{figure}[ht]
  \begin{center}
    \includegraphics[width=8.7 cm]{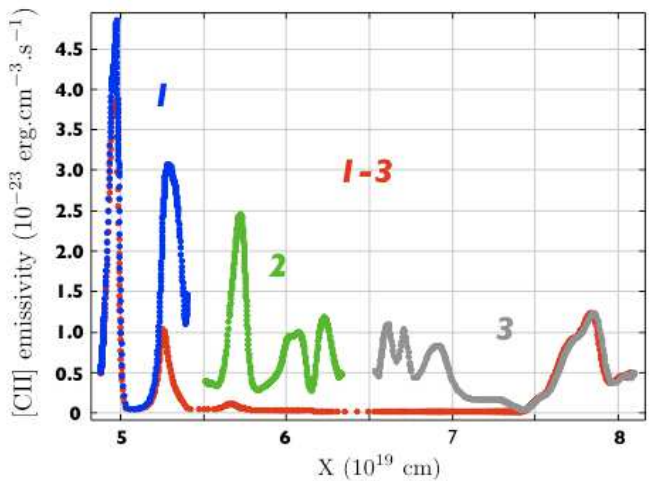}
  \end{center}
  \caption{Local emissivity of the {\sc [Cii]} 158$\mu${\rm m} line obtained from the Meudon PDR code, along the line of sight. Each plot shows the emissivity computed when using the corresponding clumps from Fig. \ref{levrierf_fig:fig3} as input density profiles to the PDR code.}  
\label{levrierf_fig:fig4}
\end{figure}

As explained before, we ran the PDR code a total of four times : once for every clump 1 to 3, and once over the entire cloud 1-3 (marked in red on Fig. \ref{levrierf_fig:fig3}). Results are summarized on Fig. \ref{levrierf_fig:fig4} and in Tab. \ref{levrierf_tab:tab1}

The effect of shielding due to the one-dimensional geometry assumed by the PDR code appears quite clearly, as the {\sc [Cii]} emission from the inner regions of the cloud is almost completely suppressed compared to results obtained when running the code on separate clumps. This is of course because photons that are energetic enough $(h\nu \geq 11.3~\mathrm{eV})$ to ionize atomic carbon are absorbed on the surface of the cloud and cannot reach these inner regions in the one-dimensional approach. 

More realistically, given the complex structures in three-dimensions, UV photons may penetrate the cloud much more deeply from other directions, and as a result, the radiation field should be higher than expected from the results of the PDR code applied to the entire cloud. 

The line being optically thin \citep[see also optical depths in Tab.~\ref{levrierf_tab:tab1}]{levrierf:N95}, we can run the code on clumps and add up contributions from these to obtain a more accurate estimate of column densities and line emissivities.

\begin{table}[bht]
  \caption{{\sc Hi} column densities, {\sc [Cii]} 158~$\mu${\rm m} integrated emissivities and line-center optical depths computed by the PDR code for clumps 1, 2, 3, as well as for clump 1-3 (marked in red on Fig. \ref{levrierf_fig:fig3}). The sums of column densities, integrated emissivities and optical depths for all three clumps are also indicated.}
  \label{levrierf_tab:tab1}
  \begin{center}
    \leavevmode
    \footnotesize
    \begin{tabular}[h]{cccc}
      \hline \\[-5pt]
      Clump \# & $N_{\sc HI}$ (cm$^{-2}$)      &  $I_{C^+}$ (erg.cm$^{-2}$.s$^{-1}$.sr$^{-1}$) & $\tau_c$\\
      \hline \\[-5pt]
      1  & 7.87~$10^{19}$ & 6.55~$10^{-6}$ & 0.39\\
      2  & 1.39~$10^{20}$ & 5.41~$10^{-6}$ & 0.34\\
      3  & 1.57~$10^{20}$ & 6.86~$10^{-6}$ & 0.99\\
      1-3  & 1.70~$10^{20}$ & 7.21~$10^{-6}$ & 1.19\\
      1+2+3 & 3.70~$10^{20}$ & 1.88~$10^{-5}$ & 1.72\\
      \hline \\
      \end{tabular}
  \end{center}
\end{table}

The data in Tab. \ref{levrierf_tab:tab1} hints at an empirical linear correlation between {\sc Hi} column density and [{\sc Cii}] 158$\mu${\rm m} integrated emissivity from the cold and dense regions of the cube, which reads
$$
\label{levrierf_eq:eq2}
\frac{\left<I_{C^+}\right>}{10^{-6}~\mathrm{ergs.cm^{-2}.s^{-1}.sr^{-1}}}=5\times\frac{\left<N_{\sc HI}\right>}{10^{20}~\mathrm{cm}^{-2}},
$$
Obviously, this is a very crude approach, as one of the five data points is actually derived from three others, and in any case this relationship may well not apply to more diffuse regions. To make a comparison with observational data, we may refer to \citet{levrierf:I02}, who give 
$$\frac{I_{C^+}}{10^{-6}~\mathrm{ergs.cm^{-2}.s^{-1}.sr^{-1}}}=0.32\times\left(\frac{I_{100}+2.58I_{60}}{\mathrm{MJy.sr}^{-1}}\right)$$
using ISO observations of high Galactic latitude clouds and IRAS far infrared data. The FIR-{\sc Hi} correlation is in turn given by \citet{levrierf:BP88}, using IRAS data and combined {\sc Hi} surveys. Averaged over mid-range Galactic latitudes $27.5^\circ<|b|<32.5^\circ$, we have
$$
\frac{I_{60}}{\mathrm{MJy.sr^{-1}}}=2.3~10^{-21}\left(\frac{N_{HI}}{\mathrm{cm}^{-2}}\right) \quad \mathrm{and}\quad \frac{I_{60}}{I_{100}}=0.18
$$
By combining the two we get
$$
\frac{I_{C^+}}{10^{-6}~\mathrm{ergs.cm^{-2}.s^{-1}.sr^{-1}}}=0.49\times\frac{N_{\sc HI}}{10^{20}~\mathrm{cm}^{-2}},
$$
which is a factor ten lower than our relation, and in agreement with COBE data \citep{levrierf:B94}. On the other hand, the {\sc [Cii]}/{\sc Hi} ratio derived from BIRT data on the Galactic {\sc Hii} region W43 yields
$$
\frac{I_{C^+}}{10^{-6}~\mathrm{ergs.cm^{-2}.s^{-1}.sr^{-1}}}=37\times\frac{N_{\sc HI}}{10^{20}~\mathrm{cm}^{-2}}.
$$
This shows that the {\sc [Cii]}/{\sc Hi} ratio strongly depends on the physical conditions, and that more work is definitely needed to understand this correlation. For the sake of simplicity, and given that our relationship happens to be a compromise between the observational data above, let us assume that we can use it to straightforwardly derive a [{\sc Cii}] emission map over the entire field. 

It remains that what we have in the data cube is the total gas content, so we need to estimate the atomic fraction, which we do crudely using a fit\footnote{To be exact, we use model 2 from \citet{levrierf:B75}, although the mean total gas density in our cube is $\left<n_{\rm tot}\right>\simeq 5$ cm$^{-3}$, which would point to model 1. The difference between the two, as far as the derived [{\sc Cii}] intensities are concerned, is negligible.} to the data in Fig. 6 of \citet{levrierf:S77}. 

\begin{figure}[ht]
  \begin{center}
    \includegraphics[width=8.7 cm]{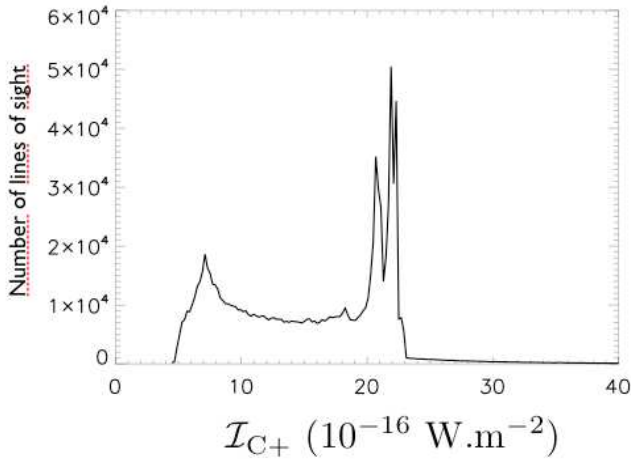}
  \end{center}
  \caption{Histogram of the [{\sc Cii}] 158$\mu${\rm m} fluxes on the entire 50 {\rm pc} $\times$ {\rm 50 pc} field, using our empirical relationship and the total gas to {\sc Hi} conversion described in the text.}  
\label{levrierf_fig:fig6}
\end{figure}

This allows us to derive a histogram of [{\sc Cii}] 158 $\mu$m line intensities emerging from the cube, using the solid angle value $\mathrm{d}\Omega$=6.1$\times 10^{-10}$~sr for the SAFARI pixels. This histogram is showed on Fig.~\ref{levrierf_fig:fig6}, and we will use it in the next section to estimate a mapping speed for the entire 50~pc~$\times$~50pc cloud.

Before that, however, we should estimate the [{\sc Cii}] emission associated with the diffuse regions of the warm ionized medium (WIM) phase, if only to check that this contribution can be neglected. To this end, we used the cosecant law given by \citet{levrierf:R92}, 
$$I^{\mathrm{WIM}}_{C^+}=3.6\times 10^{-7}\csc |b| ~\mathrm{ergs.cm^{-2}.s^{-1}.sr^{-1}}.$$
Given that there is another cosecant law for the {\sc Hi} column density \citep{levrierf:DL90},
$$\left<N_{\sc HI}\right>=(3.84\csc |b| -2.11)\times 10^{20} ~\mathrm{cm^{-2}},$$
we may combine the two to have an estimate of the WIM's contribution to [{\sc Cii}] emission,
$$
\label{levrierf_eq:eq1}
\frac{\left<I^{\mathrm{WIM}}_{C^+}\right>}{10^{-6}~\mathrm{ergs.cm^{-2}.s^{-1}.sr^{-1}}}=0.094\left(\frac{\left<N_{\sc HI}\right>}{10^{20}~\mathrm{cm}^{-2}}+2.11\right).
$$
Using the same total gas to {\sc Hi} conversion as before, this time on the mean column densities over the whole field, we find that $\left<N_{\rm tot}\right>=8.1~10^{20}$ cm$^{-2}$ implies a molecular fraction of about 15\%, so that $\left<N_{\sc HI}\right>=6.8~10^{20}$ cm$^{-2}$ and therefore $\left<I^{\mathrm{WIM}}_{C^+}\right>=8.3\times 10^{-7}$ ergs.cm$^{-2}$.s$^{-1}$.sr$^{-1}$, or $\left<\mathcal{I}^{\mathrm{WIM}}_{C^+}\right>=5.1\times 10^{-19}$ W.m$^{-2}$ in terms of emerging flux for each pixel in the map. Consequently, this contribution to the total [{\sc Cii}] emission is completely negligible.

\section{SAFARI mapping speed}
 
The focal plane array (FPA) on SAFARI's band 3 (110$\mu$m-210$\mu$m) is a 20 $\times$ 20 matrix, with a 1.9'~$\times$~1.9' field of view, so to map the entire 1.63$^\circ$ $\times$ 1.63$^\circ$ area we require $(1.63\times 60/1.9)^2 \simeq 2600$ pointings of the instrument.

The 5$-\sigma$, 1-hour sensitivity of SAFARI at 158 $\mu$m is $S_0=$1.6~10$^{-19}$ W.m$^{-2}$. Therefore, the time $\tau(\mathcal{I})$ needed to detect an emerging flux $\mathcal{I}$ is $\tau_0(S_0/\mathcal{I})^2$, with $\tau_0=5~$hours. This allows us to estimate the time needed for each of the 2600 pointings, which ranges from $\sim$1 to ~$\sim$24 seconds. By adding these up, the total time needed to map the entire sky area is $T\simeq 4.5$ hours. For the same sky area, Herschel would require some 900 hours, making such a project unrealistic.

Of course this neglects the time needed to scan the band, since in Fourier-Transform Spectrometers such as SAFARI, this is done by translating a mirror. Depending on the spectral resolution $R$ required, this can take from $\sim$ 10 s ($R=100$) to up to 3.5 min ($R=2000$). For the 2600 pointings mentioned above, this means respectively 7.2 hours and 152 hours. These are long overhead times, but one gets the full 34-210 $\mu$m band in return, which may contain many other interesting lines, such as [{\sc Nii}] at 122 $\mu$m and 206 $\mu$m.

\section{Conclusions and perspectives}

This work demonstrates the ability of SPICA/SAFARI to map large areas of the sky in the [{\sc Cii}] line in a much shorter time than what would be possible with Herschel, making it clear that the extraordinary sensitivity of the proposed mission is a definitive asset regarding this type of project.

On another level, the work presented here is part of an ongoing ASTRONET project dubbed STAR FORMAT, which is a German-French collaboration whose two main objectives are :

1/ Producing databases to publicize simulation results from MHD and PDR codes;

2/ Developing MHD, PDR and radiative transfer codes in an interoperable fashion, to allow for a much improved treatment of the physics of the ISM, especially regarding the formation of molecular clouds and dense cores.

With respect to this project, our work shows that a truly three-dimensional PDR code is a must, to adequately deal with the complex geometries of the structures formed in the MHD simulations. The development of parallel computing schemes is also quite inevitable, and it should be noted that the PDR code can already be run on a grid. It is hoped that the present paper can be used as ground work for these developments.

\end{document}